\documentstyle[12pt,epsf]{article}

\font\twelve=cmbx10 at 15pt
\font\ten=cmbx10 at 12pt

\renewcommand{\thefootnote}{\fnsymbol{footnote}}

\begin{document}

\begin{titlepage}

\begin{center}

{\ten Centre de Physique Th\'eorique\footnote{Unit\'e Propre de
Recherche 7061} - CNRS - Luminy, Case 907}

{\ten F-13288 Marseille Cedex 9 - France }

\vspace{1 cm}

{\twelve SPIN EFFECTS IN LEPTON-NUCLEON REACTIONS
\footnote{Plenary talk
presented at the Workshop on Deep Inelastic Scattering and QCD,
Paris, April 24-28, 1995 (to be published in the
Proceedings). }}

\vspace{0.3 cm}
\setcounter{footnote}{0}
\renewcommand{\thefootnote}{\arabic{footnote}}

{\bf Jacques SOFFER}

\vspace{2 cm}

{\bf Abstract}

\end{center}

We summarize some theoretical issues, which have been
considered of special importance in the discussions of
the above working group session.

\vspace{3 cm}

\noindent Number of figures : 1

\bigskip

\noindent July 1995

\noindent CPT-95/P.3210

\bigskip

\noindent anonymous ftp or gopher: cpt.univ-mrs.fr

\end{titlepage}

\section{Introduction}
 In this summary talk we would like to cover the following topics.
First we will comment on the interpretation of new results from SLAC
and SMC on measurements of the spin-dependent structure functions
$g_1(x)$ with different targets. Testing the validity of sum rules is
an important issue for perturbative QCD which is also related, in
particular, to the determination of the value of the proton spin
carried by the strange quarks $\Delta s$. There are some arguments,
based on $SU(3)$ breaking effects, which introduce uncertainties, such
that $\Delta s$ might turn out to be close to zero. The same
conclusion can be obtained by using positivity arguments and the fact
that there are very few strange quarks in the proton. Of course this
is also related to the total light quark contribution to the proton
spin $\Delta\Sigma$, whose value is now, more accurately known, larger
than before and getting closer to the naive expectation. However one
should keep in mind that all these experiments measure indeed $g_1(x)$,
not its first moment, and one should, in the first place, try to
understand properly the useful information contained in these
$x$-distributions. Next we will recall the importance of the other
polarized structure function, so-called $g_2(x)$, its origin, its
properties and what we expect to learn from its measurement which is
now under way, at the very early stage. Finally, we will review briefly
the spin programme which will be undertaken in a few years time at BNL,
by using RHIC as a polarized proton-proton collider. We will recall the
main motivations of the project and, because the center-of-mass energy
will be reaching up to $500$ GeV, it will be possible, for the first
time, to test the spin sector of perturbative QCD. We will also indicate
that it will allow to pin down, in a unique way, some polarized parton
distributions which are not directly accessible in polarized deep
inelastic scattering experiments.

\section{Spin content of the nucleon and the $g_1(x)$ structure
function}
Seven years ago, the first results of the EMC at CERN\cite{[1]} on the
$g_1^p(x)$ structure function on a proton target were rather
surprizing, because they uncovered a very serious defect in the
Ellis-Jaffe sum rule \cite{[2]}. This defect was confirmed by recent
experiments from SMC and SLAC \cite{[3]} and has been interpreted as due
to a large negative value for $\Delta s$, the contribution of the
proton spin carried by the strange quarks and a small value of
$\Delta\Sigma=\Delta u +\Delta d+\Delta s$, total contribution of the
quark spins. Such a large negative $\Delta s$ was not found by the E142
at SLAC \cite{[3]} in the measurement of the neutron structure function
$g_1^n(x)$ directly with a polarized $He^3$ target, which gives a
result consistent with $\Delta s=0$. Of course the mean value of $Q^2$
is smaller than for the earlier EMC and the more recent SMC proton
experiments, and one can be tempted to blame non-perturbative effects
for the difference between the proton and the neutron cases. One more
piece of data coming from SMC and SLAC on deuteron targets \cite{[3]},
leads also to a negative value for $\Delta s$. At this stage it is
important to recall that the experimental determination of $\Delta s$
and $\Delta\Sigma$ relies on exact $SU(3)$ flavor symmetry which is a
questionable assumption. If one ignores $SU(3)$ flavor breaking, it was
shown \cite{[4]} that there is strong correlation between $\Delta\Sigma$
and $\Delta s$, e.g. $\Delta\Sigma=0.337+0.57\Delta s$ for the proton
case, which disappears when one takes into account $SU(3)$ breaking
effects. In a recent work \cite{[5]}, one postulates that $F$ and $D$
parameters are related {\it only} to the valence quarks and the $SU(3)$
symmetry breaking for the decay $\Sigma^-\to n$ is realized by a
suppression of the strange pair production in both $n$ and $\Sigma^-$
seas and described by one parameter $\varepsilon$. In this case one can
show \cite{[4]}, by studying the dependence of $\Delta\Sigma$ and $\Delta
s$ on $\varepsilon$, both for proton and deuteron, that $\Delta\Sigma$
remains around $0.3$ and is almost insensitive to the value of
$\varepsilon$, whereas $\Delta s$ varies strongly between $-0.1$ and
$-0.02$. In a different approach \cite {[6]} where one reanalyzes hyperon
beta decay to extract the $F/D$ ratio, one is assuming that the $SU(3)$
breaking can be evaluated just in terms of mass difference of the
baryons. Instead of the value $F/D=0.575\pm 0.016$ generally used, one
finds a smaller value with a large uncertainty $F/D=0.49\pm 0.08$. In
this case there is no violation of the proton Ellis-Jaffe sum rule and
again $\Delta s$ is consistent with zero. Finally there is another
independent argument for $\Delta s$ small based on positivity
\cite{[7]}, namely one should have $|\Delta s(x)|\leq s(x)$ for all $x$.
Using the data on $s(x)$ extracted from charm production in neutrino deep
inelastic scattering, one finds that the strange quark distribution is
essentially dominated by the Pomeron which goes like $1/x$ and is spin
independent. As a result it follows that $\Delta s$ is consistent with
zero, but perhaps this analysis should be reconsidered by using more
recent CCFR data \cite{[8]}. To conclude this discussion we think it
would be extremely useful to have a direct measurement of $\Delta s$
and any suggestion is very welcome\footnote{$\Delta s$ can also be
measured in elastic $\nu p$ scattering by separating the axial form
factor contribution $G_A(Q^2)$ extrapolated to $Q^2=0$. The existing
data \cite{[9]} yields a large negative $\Delta s$ with large errors
which are due to several uncertainties in getting $G_A(0)$\cite{[10]}.}.


\def\epsfsize#1#2{0.7#1}

\epsfverbosetrue

\vspace{0,6cm}

\hspace{0.3cm}
\epsfbox{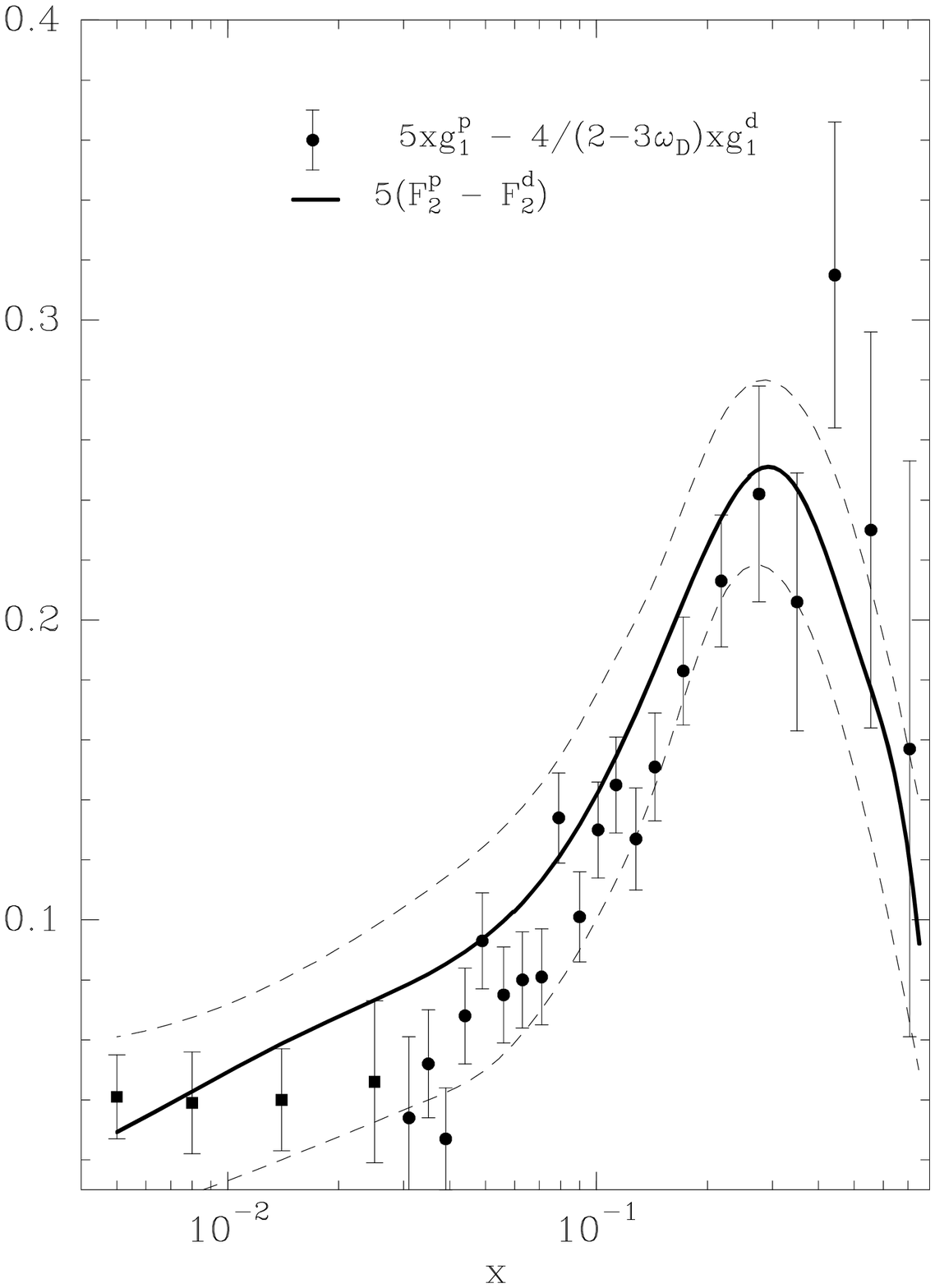}

\vspace{0,5cm}

\centerline{Fig. 1 - Experimental test of eq.(2) (see text) taken
from ref.[13]}
\vskip 1truecm

As indicated before, we believe that the $x$-dependence of the $g_1(x)$
structure functions contains relevant information which is somehow
washed out when one considers only first moment sum rules\cite{[11]}.
Moreover testing sum rules involves necessarily some uncertain
extrapolations of the data due to the limited kinematic range
accessible in any experiment. We will now briefly illustrate this point
as follows. The $g_1(x)$ structure functions for proton, neutron and
deuteron are expressed in terms of three distributions $\Delta u(x),
\Delta d(x),\Delta s(x)$ (here $u$ means $u+\bar u$ etc...). It is well
known that $g_1^p(x)-g_1^n(x)$ is independent of $\Delta s(x)$ and
similarly one can eliminate $\Delta d(x)$ by considering
$4g^p_1(x)-g^n_1(x)$. Let us now assume $\Delta s(x)\equiv 0$ and, as
explained in Ref.\cite{[12]}, let us postulate the following simple
relation between unpolarized and polarized distributions

\begin{equation}
\Delta u(x)=u(x)-d(x)\ .
\end{equation}

It implies
\begin{equation}
5xg^p_1(x)-4/(2-3\omega_D)xg^d_1(x)=5(F_2^p(x)-F_2^d(x))
\end{equation}
a simple relationship between two $g_1$ and two unpolarized structure
functions $F_2(x)$ directly measured by experiment. We have used the
standard relations between deuteron, proton and neutron and for the
polarized case and $\omega_D$ is the $D$-state probability in the
deuteron. We show in Fig.1 an experimental test of eq.(2). We have
tested eq.(2) by using for the l.h.s., the


\noindent SLAC data on the $g_1$'s at
$Q^2=3GeV^2$ (full circles) and
for the r.h.s., the NMC parametrization
for $F^p_2$ and $F^d_2$ (full line and dash lines for the estimated
errors)\cite{[14]}. In the small $x$ region we have also included the
preliminary SMC data\cite{[3]} (full squares). The test is indeed very
well satisfied and gives, within the present experimental errors, a
fairly good support to eq.(1) and $\Delta s(x)\equiv 0$. Moreover if
one takes the first moment of both sides of eq.(2) using also
ref.\cite{[15]}, one finds for the l.h.s. $0.588\pm 0.054$ and for the
r.h.s. $0.587\pm 0.065$ which are in remarkable agreement.

Let us end by making a few remarks on the prospects of the $g_1$
structure functions. First in testing the sum rules, one is relying on
the crucial assumption that the measured asymmetry $A_1(x,Q^2)$ is
independent of $Q^2$, which allows to rescale the $g_1(x,Q^2)$ obtained
at different $Q^2$, to a single $Q^2$ value. This has to be more firmly
established. Second, the Bjorken sum rule is now satisfied up to
13\%\cite{[3]} and this important test must be confirmed at a higher
level of accuracy. Of course this is connected to the behaviour of
$g_1(x)$ in the very small $x$ region which seems to be different for
proton and neutron according to the recent SMC data\cite{[3]}. This
puzzling situation must be resolved urgently.

\section{The $g_2$ structure function}

In deep inelastic $\mu(e)$ scattering with the lepton beam
longitudinally polarized and the target with the spin transverse with
respect to the beam direction, one can measure a spin asymmetry which
is related to a ''transverse'' spin-dependent structure function
$g_T(x,Q^2)$. It turns out that $g_T$ has a simple expression in terms
of $g_1$ and another structure function, so called $g_2(x,Q^2)$, since
we have $g_T=g_1 + g_2$. The basic properties of $g_2$ have been nicely
summarized in the contribution of X. Ji\cite{[16]}. In the simple
parton model, as a consequence of helicity conservation, one finds that
$g_T$ must vanish and therefore, in the scaling limit, one has
$g_2(x)=-g_1(x)$. However from an analysis in terms of operator product
expansion (OPE) on the light cone, one finds that the situation is not
so simple and $g_2$ can be decomposed into two sets of operators. The
first set is twist-2 operators, which are the same as those of the
decomposition of $g_1$. The second set is twist-3 operators, which
involve quark-gluon correlation functions. Therefore one can write

\begin{equation}
g_2(x,Q^2)=g_2^{WW}(x,Q^2)+\bar g_2(x,Q^2)\ ,
\end{equation}
where $g^{WW}_2$ and $\bar g_2$ correspond to twist-2 and twist-3
respectively. It was shown that $g^{WW}_2$ is fully determined by $g_1$
because one has\cite{[17]}

\begin{equation}
g_2^{WW}(x,Q^2)=-g_1(x,Q^2)+\int^{1}_{x}\frac{dy}{y}g_1(x,Q^2),
\end{equation}
but since there is a priori no theoretical reason to  expect $\bar
g_2<<g^{WW}_2$, it is important to measure $g_2$.

Very preliminary results were obtained recently both on proton and
deuteron targets at SLAC\cite{[3]} and, although the errors are large,
they seem to indicate that for proton, $\bar g_2$ is small except in a
region around $x=0.01$. There is also the Burkhardt-Cottingham (BC) sum
rule\cite{[18]} which says that
\begin{equation}
\int^{1}_{0}dx g_2(x,Q^2)=0
\end{equation}
and this important result has to be checked experimentally. The $Q^2$
dependence of the BC sum rule and its possible violation in the very
low $Q^2$ region is an interesting problem which has new
implications\cite{[19]}. In perturbative QCD, the validity of the BC
sum rule was checked\cite{[20]} at order $\alpha_s$ on a quark target
of mass $m$ to all orders in $m^2/Q^2$, a result which might help to
clarify the nucleon target case. Finally, it is worth recalling the
relevance of the second moment of $g_2$ and, in particular, the twist-3
coefficient $d^{(2)}$ for which there are different theoretical
predictions\cite{[16]} which will have to be confronted to future
accurate data.

\section{Spin physics at RHIC}

Before one can come up with a realistic picture of the nucleon, many
fundamental questions remain to be answered and, in particular as we
have seen above, in the area of the polarized parton distributions.
Polarized deep inelastic scattering provides some valuable insight in
this direction, but polarized hadron-hadron collisions at high energy
give access to new spin-dependent observables which contain also, in
some cases, a far more unique information. A Relativistic Heavy Ion
Collider (RHIC) is now under construction at Brookhaven National
Laboratory and more than four years ago, it was already realized that
one should propose a very exciting physics programme, provided this
machine could be ever used as a polarized $pp$ collider. Of course all
these considerations relie on the foreseen keys parameters of this new
facility, i.e. a luminosity up to $2.10^{32} cm^{-2} sec^{-1}$ and an
energy of $50-250$ GeV per beam with a polarization, either
longitudinal or transverse, of $70\%$. Since then, the RHIC Spin
Collaboration (RSC) has produced a letter of intent\cite{[21]} and has
undertaken several serious studies in various areas, leading to a
proposal\cite{[22]} which is now fully approved. Both detectors STAR and
PHENIX of the heavy ion programme will be involved and the first data
taking is expected by 1999. Some of the physics topics which will be
explored at RHIC have been reviewed in the working group by A.
Sch\"afer\cite{[23]} who also presented many interesting results from a
dedicated Monte Carlo code. The magnitude and the sign of the polarized
gluon distribution $\Delta G(x,Q^2)$ has to be determined because it is
believed to affect the quark helicity distributions $\Delta q(x,Q^2)$,
via the axial anomaly\cite{[24]}, and it is also needed to perform
their $Q^2$ evolution. So far nothing is known about $\Delta G(x,Q^2)$
experimentally, but it can be measured by the double helicity asymmetry
$A_{LL}$, for example, in direct $\gamma$ production and in jet
production. Direct photon production in $pp$ collisions is largely
dominated, at leading order, by the Compton diagram $qg\to q\gamma$ and
therefore $A_{LL}$ provides an almost direct measurement of $\Delta
G(x,Q^2)/G(x,Q^2)$. The number of subprocess contributing to jet
production is larger, because we have $gg$, $gq$ and $qq$ as initial
states. However for one-jet production, gluon-gluon scattering
dominates the $p_T$ spectrum in the low $p_T$ region, i.e. for
$10<p_T<20$ GeV/c at the RHIC energy, and this also allows to extract
$\Delta G$. In the large $p_T$ region (i.e. $p_T>50$ GeV/c), the cross
section is dominated by quark-quark scattering, so $A_{LL}$ will be
driven by $(\Delta u/u)^2$. Another very efficient way to isolate
$\Delta q/q$ or $\Delta \bar q/\bar q$ for both $u$ and $d$ quarks, is
by measuring the single (or double) helicity parity-violating asymmetry
in weak boson production\cite{[25]}. This is due to the anticipated
high luminosity (i.e. $800 pb^{-1}$ in three months running time) which
will allow to collect a very copious number of $W^{\pm}$ and $Z$
bosons. In addition the polarized antiquark distribution can be very
well measured through $A_{LL}$ in lepton pair production, described by
the standard Drell-Yan mechanism and, for example, if $\Delta\bar q=0$
one will observe $A_{LL}=0$.

Finally, it is worth mentioning that with the option of transversely
polarized protons at RHIC, we will be able to measure directly, for the
first time, the leading twist-2 quark transversity distribution, so
called $h_1^q(x)$\cite{[26]}. Clearly there is one such distribution for
each flavor ($u,d,$ etc...) and for both quark and antiquark, but
nothing is known experimentally about them. At the level of the parton
model, there is a useful bound\cite{[27]}, implied by positivity, which
reads

\begin{equation}
q(x)+\Delta q(x)\geq 2|h^q_1(x)|\ .
\end{equation}

Once more, lepton pair production and $Z$ production are the best
hadronic probes for these new transversity distributions which are
simply related to the double transverse spin asymmetry
$A_{TT}$\cite{[12]}.

\begin{center}
{\bf Acknowledgements}
\end{center}

I am glad to thank the organizers of this Workshop, in particular
Madame V. Brisson. It was a pleasure to set up the programme of the
Working Group III with E. Hughes and I also thank X. Ji, J.
Lichtenstadt, Ph. Ratcliffe and A. Sch\"afer for their contributions.

\end{document}